\documentclass[preprint,aps,pra,showpacs]{revtex4}
\usepackage{tabularx}
\usepackage{dcolumn}
\usepackage{graphics}
\usepackage{bm}
\usepackage[dvips]{graphicx}
\usepackage[dvips]{rotating}
\usepackage[latin1]{inputenc}
\usepackage{dcolumn}
\usepackage{tabularx}
\usepackage{amsmath}
\usepackage{amssymb}
\usepackage{float}
\begin{document}
\draft

\title{Photodetachment near a repulsive center}

\author{B. C. Yang$^1$, J. B. Delos$^2$ and M. L. Du$^1$}
\email{duml@itp.ac.cn}

\affiliation{$^1$State Key Laboratory of Theoretical Physics,Institute of Theoretical Physics, Chinese Academy of Science, Beijing 100190, China\\
$^2$Physics Department, College of William and Mary, Williamsburg, Virginia 23185, USA}

\date{\today}

\begin{abstract}
We study the total photodetachment cross section of H$^-$ near a repulsive center. An analytical formula for the photodetachment cross section is obtained
using the standard closed-orbit theory and extending it to the energy range below the zero-field threshold. The formula is found to be accurate by
comparing with an exact quantum calculation. A comparison with the photodetachment cross section in an effective homogeneous electric field is made, and we discuss
the similarities and differences of the two systems.

\par


\end{abstract}

\maketitle

\section{Introduction}

In the past three decades, many theoretical efforts have been devoted to reveal and investigate the rich and novel phenomena in the photodetachment
spectra of negative ions in external fields, among which the closed-orbit theory developed around 1987\cite{COT} supplies a general semiclassical
picture for the external-field effects (note that a similar idea can also be found in Ref.\cite{Fabrikant} ). Other more accurate but much more
complicated considerations have been made which also have achieved great success\cite{FT, Fabrikant_exact, QQES, Bracher_exact, Bracher_thesis,
Bracher}. Both the predicted oscillatory structures above threshold and the quantum tunneling effect below threshold have been observed clearly in
the related experiments\cite{experiment01, experiment02}.

Very recently, based on our previous studies on the photodetachment of negative ion in an external field, and stimulated by an experiment of the
Washington group \cite{washington} where they reported the observed angle-resolved photoelectron spectra from doubly charged anions having the
structure of a linear chain, $^-O_2C-(CH_2)_m-CO_2^-$ (with $3\leq m\leq11$), we analyzed in detail the spatial interference structures on a screen
for a theoretical model of negative-ion photodetachment near a repulsive center\cite{BCYang1}. In this case, a unique closed orbit can be found on
the line between the electron source and the repulsive-force center.

In the present paper, we use a well-established H$^-$ model to investigate the effect in the photodetachment spectra induced by this returning closed
orbit. After briefly depicting the theoretical model and the electron-wave propagation after detachment in Sec. \textrm{II}, we first use the
standard closed-orbit theory\cite{COT, Du1} to obtain a formula for the oscillatory structure in the photodetachment spectra above the zero-field
threshold. Then we extend it naturally to the energy range below the zero-field threshold by introducing a ``uniform approximation''
(Sec. \textrm{III}). To understand the oscillations of the photodetachment spectra better, a comparison with the photodetachment spectra in a
homogeneous electric field is made in Sec. \textrm{IV}. A brief conclusion is given in Sec.\textrm{VI}. Atomic units are used throughout this work
unless specified otherwise.

\section{Theoretical model and photoelectron wave propagation}

Here, we set the detached-electron source to be the origin of coordinate frame, and the $z$-axis is along the direction from the repulsive-force
center to the electron source (Fig. 1). To distinguish the new frame from the previous one in Ref.\cite{BCYang1}, we use a prime to denote the
coordinates ($r'$, $\theta'$) relative to the repulsive center.

The zero-field photodetachment of H$^-$ has been well established in the literature\cite{Du2, Du3, Ohmura}. Here we only sketch the theoretical model
briefly. Initially the electron is loosely bound by a short range potential and the initial bound state can be written as
\begin{equation}\label{1}
    \psi_i = B\frac{e^{-k_br}}{r}~,
\end{equation}
where $B=0.31552$ is a normalization constant and $k_b=\sqrt{2E_b}$ with the binding energy $E_b=0.7542eV$. After irradiation by laser light linearly
polarized along the $z$-direction, the generated directly-outgoing wave is given by
\begin{equation}\label{2}
    \psi_{out} = \frac{4Bki}{(k_b^2+k^2)^2}\frac{e^{ikR}}{R}\cos\beta~,
\end{equation}
where $\beta$ is the initial ejection angle of the photodetached electron, and $k=\sqrt{2E}$ with $E$ denoting the initial kinetic energy. $R$
satisfies $\sqrt{\frac{d}{2\alpha \widetilde{E}}}\ll R \ll \frac{\widetilde{E}}{\widetilde{E}+1}d$. Actually, one can find the following calculation
is independent of such a surface $R_s$.

After being detached from the loosely bound initial state, the detached-electron motion is governed by the following Hamiltonian
\begin{equation}\label{3}
    H=\frac{p_{r'}^2}{2}+\frac{p^2_{\theta'}}{2r'^2}+\frac{\alpha}{r'}-\frac{\alpha}{d}~,
\end{equation}
where $d$ is the distance between the wave source and the repulsive center, and $\alpha$ denotes the negative-charge number at the force center. The
classical trajectory analysis for the present model has been presented in our previous work\cite{BCYang1}, from which one can immediately find a
unique closed orbit lying on the line between electron source and the repulsive-force center (see Fig. 2 in Ref. \cite{BCYang1}).

All the photoelectron wave propagates away from the source region to infinity except a small part of the outgoing wave that returns to the source
region along the closed orbit. Semiclassical propagation of the detached-electron wave along the closed orbit gives the returning wave as,
\begin{equation}\label{4}
    \psi_{ret}=\psi_{out}Ae^{i(S-\mu\frac{\pi}{2})}~,
\end{equation}
where the Maslov index $\mu$ equals one as this closed orbit touches a caustic surface just once at the turning point; $A$ and $S$ are, respectively,
the semiclassical amplitude and the accumulated phase (action variation) of an associated wave propagating along the closed orbit.

When the detached-electron wave comes back to the source after semiclassical propagation along closed orbit, from Eqs.(23) and (26) in
Ref.\cite{BCYang1}, we get the following expression for the semiclassical amplitude,
\begin{equation}\label{5}
    A=\frac{R}{d}\cdot\frac{1}{\big|\big(\frac{\partial\theta'}{\partial\beta}\big)_{r'=d}\big|}~,
\end{equation}
with,
\begin{equation}\label{6}
    \bigg(\frac{\partial\theta'}{\partial\beta}\bigg)_{r'=d}=-4\widetilde{E}~,
\end{equation}
where $R$ is an initial spherical surface around the source, and $\widetilde{E}=Ed/\alpha$ is the defined scaled energy as before. Hence, we get the
amplitude for the returning wave as,
\begin{equation}\label{7}
    A=\frac{\alpha R}{4Ed^2}~.
\end{equation}

The classical action variation along the closed orbit can be directly calculated using its definition,
\begin{equation}\label{8}
    S=2\int^d_{r_{turn}}\sqrt{2\Big(E+\frac{\alpha}{d}-\frac{\alpha}{r'}\Big)}dr' ~,
\end{equation}
with the turning point $r'_{turn}$ expressed as
\begin{equation}\label{9}
    r'_{turn}=\frac{\alpha}{E+\frac{\alpha}{d}}~.
\end{equation}
By assuming $\sqrt{r'}=x$ and integrating by parts, we have,
\begin{equation}\label{10}
    S=\frac{2\sqrt{2d\alpha}}{\sqrt{1+\widetilde{E}}}\bigg[\sqrt{\widetilde{E}(1+\widetilde{E})}-\ln\Big(\sqrt{1+\widetilde{E}}+\sqrt{\widetilde{E}}~\Big)\bigg]~,~~~~~~\widetilde{E}\geq0~,
\end{equation}
which can be proved to be identical with the previous expression in Eq.(67) in \cite{BCYang1}.

The returning wave propagating along the closed orbit can be obtained by substituting the above Eqs.(2), (7) and (10) into Eq.(4). Then, after a
further connecting procedure on a bound surface $R$ as usual\cite{Du1}, we get the returning wave near the wave source as,
\begin{equation}\label{11}
    \psi^s_{ret}=-\frac{4Bki}{(k_b^2+k^2)^2}\frac{A}{R}\exp\big[i\big(kr\cos\theta+S-\frac{\pi}{2}\big)\big]~.
\end{equation}
The returning wave in Eq.(11) will overlap and interfere with the wave source, and oscillatory structures will be observed in the total
photodetachment cross section described quantitatively in the next section.

\section{photodetachment cross section}

In this section, we use the standard procedure of the closed-orbit theory to derive a cross section formula above the zero-field threshold, and then
we extend the obtained expression to a uniform one which is applicable for the energy range very close to the zero-field threshold and even below
threshold.

\subsection{Modulations above zero-field threshold induced by closed orbit}

According the closed-orbit theory\cite{COT, Du1}, the oscillatory term in the photodetachment spectra contributed by the closed orbit is given by
\begin{equation}\label{12}
    \sigma_r=-\frac{4\pi E_{ph}}{c}\Im\langle D\psi_i|\psi^s_{ret}\rangle~,
\end{equation}
where $E_{ph}= E_b+E$ denotes the absorbed photon energy; $c$ and $D$ denote the light speed and the dipole operator, respectively.

By using the dipole operator (here, $D=r\cos\theta$), the initial bound state in Eq.(1) and the returning wave in Eq.(11), the integration in Eq.(12)
can be written out as,
\begin{equation}\label{13}
    \langle D\psi_i|\psi^s_{ret}\rangle=-\frac{8\pi B^2ki}{(k_b^2+k^2)^2}\frac{A}{R}\exp\big[i\big(S-\frac{\pi}{2}\big)\big]\int^\infty_0dr\Big[r^2e^{-k_br}\int^{1}_{-1}\tau e^{ikr\tau}
    d\tau\Big]~,
\end{equation}
where, $\tau=\cos\theta$. And the integration in Eq.(13) can be worked out straightforward as
\begin{equation}\label{14}
    \int^\infty_0dr\Big[r^2e^{-k_br}\int^{1}_{-1}\tau e^{ikr\tau}d\tau\Big]=\frac{4ki}{(k_b^2+k^2)^2}~.
\end{equation}

From the above Eqs.(12)-(14), the oscillatory term induced by the closed orbit can be obtained as
\begin{equation}\label{15}
    \sigma_r=\frac{16\pi^2B^2E}{c(E_b+E)^3}\frac{A}{R}\cos(S)~.
\end{equation}
Combining the smooth background without any external field\cite{Du1, Ohmura}, the total photodetachment cross section for the simple H$^-$ model near
a repulsive coulomb center can be written out completely as,
\begin{equation}\label{16}
    \sigma=\frac{16\pi^2\sqrt{2}B^2E^{3/2}}{3c(E_b+E)^3}+\frac{4\pi^2B^2}{c(E_b+E)^3}\cdot\frac{\alpha}{d^2}\cdot\cos(S)~,~~~~\widetilde{E}\geq0~,
\end{equation}
where we substituted $A$ by Eq.(7).

\subsection{Modifications close to the threshold and extending to the tunneling region}

Very close to the zero-field threshold, the cross-section expression in Eq.(16)  is not appropriate. Besides, slightly below the zero-field
threshold, the loosely-bound electron can also tunnel out through the repulsive potential, for which a consistent formula is also needed.

First, we transform the formula in Eq.(16) into a product of the zero-field cross section and a field-induced modulation function as following
\begin{equation}\label{17}
    \sigma=\frac{16\pi^2\sqrt{2}B^2E^{3/2}}{3c(E_b+E)^3}\cdot\mathcal{H}^{F}~,
\end{equation}
and separate the modulation function into
\begin{equation}\label{18}
    \mathcal{H}^{F}=\mathcal{L}^{F}(S)+\Big(\frac{A}{kR}-\frac{1}{3S}\Big)\cdot3\cos(S)~,
\end{equation}
where
\begin{equation}\label{19}
    \mathcal{L}^{F}(S)=1+\frac{1}{3S}\cdot3\cos(S)~.
\end{equation}
Using the semiclassical amplitude in Eq.(7) and the action variation in Eq.(10), the following limiting behavior can be found when the zero-field
threshold is approached:
\begin{equation}\label{20}
    \lim_{E\rightarrow0}\bigg[\frac{A}{kR}\bigg/\frac{1}{3S}\bigg]=1~.
\end{equation}
Hence in the modulation function only the first part $\mathcal{L}^{F}(S)$ will survive.

As the photon energy is approaching the zero-field threshold, the closed orbit mentioned above will be shorter and shorter, and the pencil of the
detached-electron wave propagating near this closed orbit will be mainly affected by the repulsive potential very close to the wave source.
Therefore, given that the spatial gradient of repulsive potential is proportional to $1/r'^2$, a linear approximation is possible for the repulsive
potential very close to the zero-field threshold, and the limiting behavior of the surviving modulation function $\mathcal{L}^{F}(S)$ should be
similar with that induced by a uniform electric field.

By carefully examining the photodetachment process in a uniform electric field, we find the amplitude $A_{u}$ and the action phase $S_{u}$ along
closed orbit has a very simple but interesting relation,
\begin{equation}\label{21}
    \frac{A_{u}}{kR}=\frac{1}{3S_{u}}~.
\end{equation}
Once substituting the above equation into Eq.(18), one can immediately find that the modulation function $\mathcal{H}^{UF}(S)$ becomes identical to
$\mathcal{L}^{F}(S)$ in this limit. On the other hand, for the photodetachment of negative ion in a uniform electric field, several quantum
cross-section expressions have been obtained by Fabrikant et. al.\cite{Fabrikant_exact}, and an completely analytic $\delta$-source model has also
been well established by Bracher et. al using a quantum-source theory\cite{Bracher_exact, Bracher_thesis}, from which an exact form of the modulation
function can be abstracted.

Accordingly, based on the above considerations, we replace $\mathcal{L}^{F}$ by an exact form of the modulation function $\mathcal{H}^{UF}$ in the
uniform-field case but with a different action variation $S$, which we call ``uniform approximation''. By this the modulation function
in Eq.(18) can be extended as
\begin{equation}\label{22}
    \mathcal{H}^{F}=\mathcal{H}^{UF}(\zeta)-\Big(\frac{A}{kR}-\frac{1}{3S}\Big)\cdot6\pi Ai(\zeta)Ai'(\zeta)~,
\end{equation}
where $\zeta$ is a \emph{real variable} defined as
\begin{equation}\label{23}
    \zeta=-\Big(\frac{3}{4} S\Big)^{2/3}~,
\end{equation}
and
\begin{equation}\label{24}
    \mathcal{H}^{UF}(\zeta)=\frac{\pi}{(-\zeta)^{\frac{3}{2}}}\big[\zeta^2Ai^2(\zeta)-2Ai(\zeta)Ai'(\zeta)-\zeta Ai'^2(\zeta)\big]~.
\end{equation}
For the uniform modulation function in Eq.(22), the second part in Eq.(18) has also been uniformized using the Airy function and its derivative,
similar to what we did in Ref. \cite{BCYang1}.

The semiclassical-propagation amplitude $A$ in Eq.(7) can be extended into the negative electron-energy range directly. And the quantities of both
the electron momentum $k$ and the semiclassical-propagation phase $S$ should also be extended naturally from their original definitions (see Eq.(8)
for the original definition of $S$). Accordingly, for the photon energy below the zero-field threshold, their corresponding expressions can be
written out as
\begin{equation}\label{25}
    k=i\sqrt{2(-E)}
\end{equation}
and
\begin{equation}\label{26}
    S=-2i\int^{r'_{_{turn}}}_d\sqrt{2\Big(\frac{\alpha}{r'}-\frac{\alpha}{d}-E\Big)}dr' ~,
\end{equation}
Similar to the derivation of classical action $S$ above threshold in Sec. \textrm{III}, the above integration can be worked out as
\begin{equation}\label{27}
    S=\frac{2i\sqrt{2d\alpha}}{\sqrt{1+\widetilde{E}}}\bigg[\sqrt{-\widetilde{E}(1+\widetilde{E})}+\arcsin\Big(\sqrt{1+\widetilde{E}}~\Big)-\frac{\pi}{2}\bigg]~,~~~~~~-1<\widetilde{E}\leq0~,
\end{equation}
and
\begin{equation}\label{28}
    S=-i\infty~,~~~~~~\widetilde{E}\leq-1~.
\end{equation}

In Fig.2, we display the behavior of the photodetachment spectra for H$^-$ near a repulsive center using a representative case with $d=200a_0$ and
$\alpha=1$, where both the oscillatory structure above the zero-field threshold and the quantum tunneling effect below threshold can be observed
clearly. An exact quantum calculation based on the Coulomb Green's function is also displayed in Fig.2, which confirms the accuracy of the present formula.
The exact quantum treatment will be presented in a separate paper\cite{next-paper}.

Finally we discuss the shift of photo-electron threshold caused by the repulsive center. According to the modulation function in Eq.(22), the action value in Eq.(28) indicates that the photodetachment cross section vanishes completely once the scaled energy $\widetilde{E}$ is smaller than minus one. Therefore, the present theory suggests that, as a result of the presence of the repulsive center, the photo-electron threshold is actually shifted to a lower value 
($E_b-\alpha/d$) relative to the zero-field photodetachment threshold. This theoretical result is consistent with the observations reported by the Washington group in their experiments\cite{washington, washington02}.

\section{comparing with the photodetachment spectra in a homogeneous electric field}

By comparing the cross-section formula in Eq.(16) with that in a uniform electric field\cite{Du1}, an interesting phenomenon can be observed:
\emph{the modulation amplitude in the photodetachment spectra induced by a nearby repulsive center is identical to that induced by an effective
uniform static field strength of $F=\alpha/d^2$}, though the phase is different. The equivalence of oscillatory amplitudes in the total cross section
might be a surprise at first glance, because the differential cross sections are completely different in Ref.\cite{BCYang1}. To examine this further, we compare in Fig.3 the total cross section for the photodetachment of hydrogen negative ion near a repulsive center with that in a uniform electric field. Indeed, we can observe the same magnitude of the oscillations for the two different systems, and the physical reason can be found in the following.

Given the generality of the derivation presented in this work, the formulas in Eqs.(15), (17) and (22) also represent the oscillatory behavior of the
photodetachment spectra in a uniform electric field but with amplitude $A_{u}$ and phase $S_{u}$ for the returning wave along the closed orbit.
Therefore, to compare the two systems, we examine the related quantities in the photodetachment spectra which are, respectively, $A$ and $S$ for the
photodetachment near a repulsive center and $A_{u}$ and $S_{u}$ for the photodetachment in a homogeneous electric field.

Let us first examine the phase. For photodetachment in a uniform field $F = \frac{\alpha}{d^2}$, the modulation phase can be worked out as,
\begin{equation}\label{29}
    S_{u}=\frac{4\sqrt{2d\alpha}}{3}\widetilde{E}^{3/2}~,
\end{equation}
which is compared in Fig.4(a) with the phase $S$ in Eqs. (10) and (27) by using $d = 200a_0$ and $\alpha = 1$ for example. Actually, from Eq.(29) one
can find the ratio of $S$ to $S_{u}$ depends only on the scaled energy (Fig.4(b)). In Fig.4 we can find that only close enough to the threshold
energy, the two phases are approximately equal; As the energy increases from the zero-field threshold, the ratio monotonically decreases from unity.
At any scaled energy above threshold, the phase accumulation along the closed orbit in a repulsive force field is always smaller than that in the
corresponding uniform static field. Both the above two points are revealed clearly in Fig.3.

For the amplitudes $A$ and $A_{u}$, one can easily find the following relationship
\begin{equation}\label{30}
    \frac{A}{kR}=\frac{A_{u}}{kR}=\frac{1}{3S_{u}}~,
\end{equation}
which suggests the following two points. (1) The ratio of $S$ to $S_{u}$ displayed in Fig.4(b) is also the ratio of $\frac{A}{kR}$ to $\frac{1}{3S}$
for the photodetachment near a repulsive center, which measures the contribution of the second part in Eq.(22) when the photon energy is away from
the zero-field threshold. (2) The amplitude $A$ for the photodetachment near a repulsive center is exactly equal to the amplitude $A_{u}$ for the
photodetachment in a homogeneous electric field, which is the reason why the oscillation amplitude of the photodetachment spectra in Fig.3 is
identical.

To see this more clearly from a dynamical view, we show in Fig. 5 a pencil of trajectories propagating alongside the closed orbit. The paths with
turning points near $z=-80a_0$ (red online) are for a repulsive coulomb field, while those with turning points near $z=-150a_0$ (green online) are
for the corresponding constant electric field.  When the trajectories go out and then back to the dashed curve (blue online), each ``red'' curve
going out from the source at angle $\beta$ crosses the ``green'' curve that went out at that same angle.  Therefore the flux densities through
surfaces perpendicular to the trajectories must be equal, and those surfaces coincide at the source.

\section{Conclusion}

Based on the preceding work\cite{BCYang1}, we have explored the total photodetachment cross section near a repulsive center by using a
well-established H$^-$ model. Using the standard procedure in the closed-orbit theory, we derived a cross-section formula for the oscillatory
structures in the photodetachment spectra above the zero-field threshold. And then, by introducing a ``uniform approximation'', the
cross-section expression has been extended to a uniform formula which is applicable for the energy range below the zero-field threshold. By comparing
with the familiar case of photodetachment in a homogeneous electric field, we find the oscillatory amplitudes for the two different systems are
identical, and the ratio of their oscillatory phase is only dependent on a scaled energy.

\begin{center}
{\bf ACKNOWLEDGMENTS}
\end{center}
\vskip8pt B. C. Y. and M. L. D. acknowledge the support from NSFC Grant No. 11074260 and 11121403. J. B. D. acknowledges support from NSF Grant No.
1068344.


\newpage
\begin{figure}[H]
\centering
  \includegraphics[width=350pt]{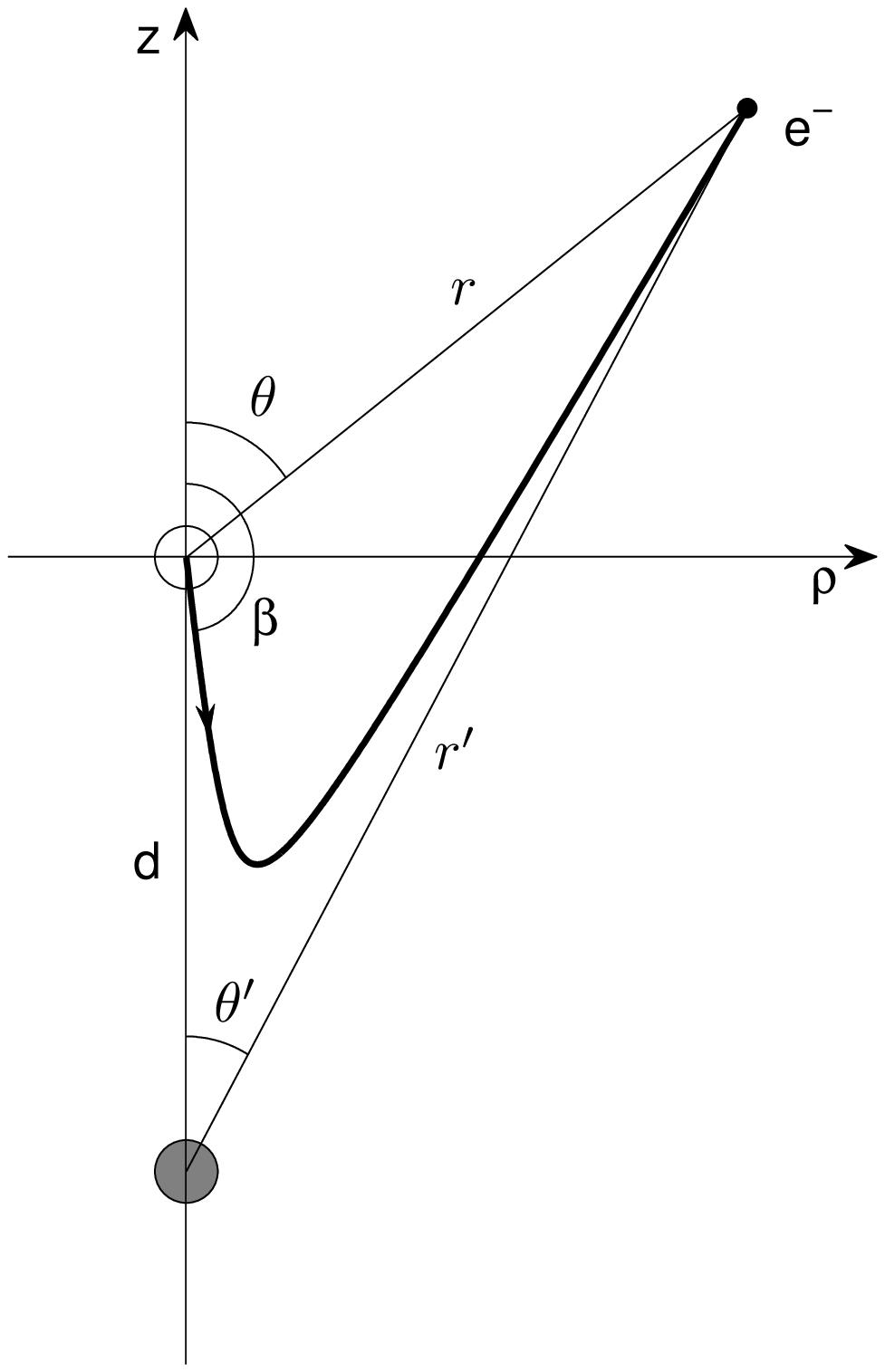}
  \caption{Schematic representation of the theoretical model and the choices of the coordinate frames.
   The open and the solid circles denote the electron source and the negatively charged center, respectively. The electron
source is supplied by the photodetachment of hydrogen negative ion. After the electron is detached by a photon, it escapes following a hyperbolic
trajectory as shown by the heavy solid curve on which a dark point denotes the classically-moving electron. There are two coordinate frames displayed
here, and we distinguish them by adding a prime to the coordinates relative to the repulsive center. The angle $\beta$ shows the initial ejection
angle from the electron source.}
\end{figure}

\begin{figure}[H]
\centering
  \includegraphics[width=350pt]{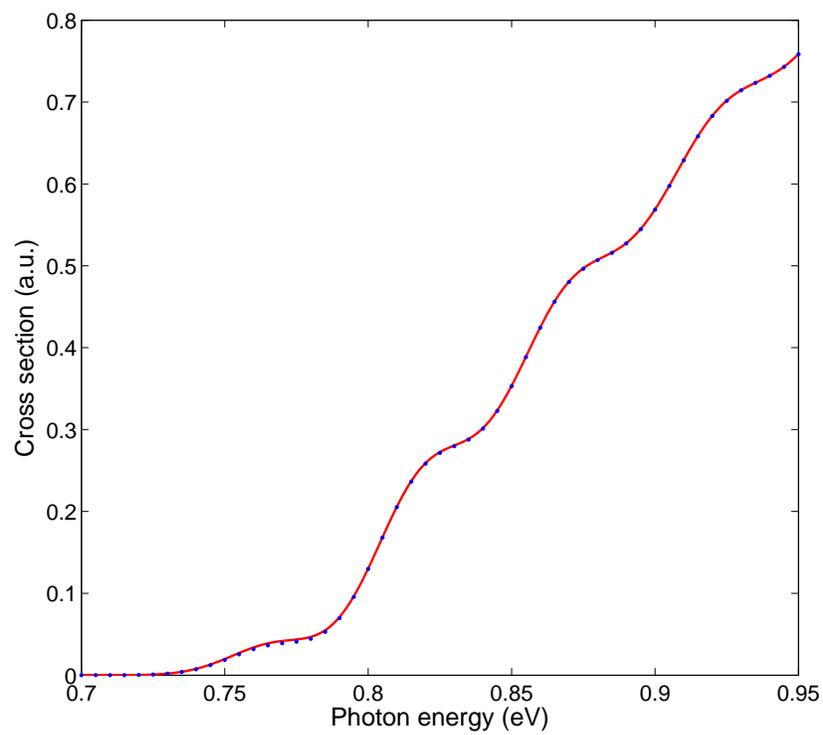}
  \caption{The photodetachment cross section near a repulsive center
  is calculated using $d=200a_0$ and $\alpha=1$. The solid curve is calculated using Eqs. (17), (22) and (24), and the exact quantum calculation is displayed by the dots.}
\end{figure}

\begin{figure}[H]
\centering
  \includegraphics[width=400pt]{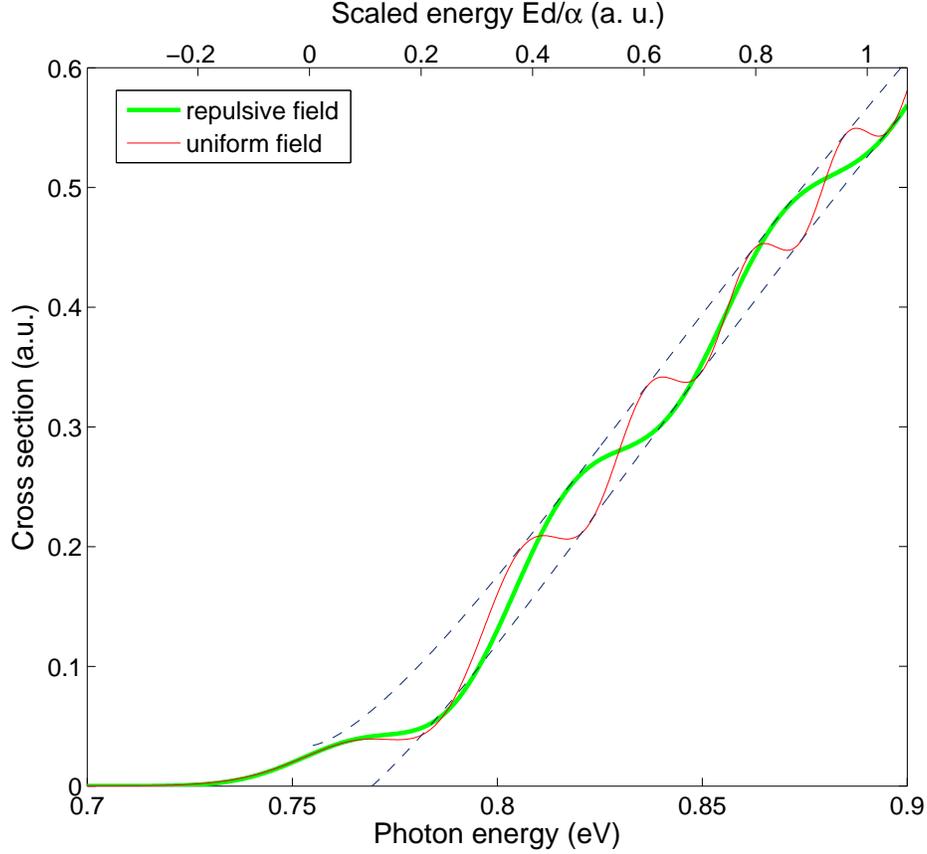}
  \caption{(Color online) Comparison of the total cross section for the photodetachment of hydrogen negative ion near a repulsive center(green and thick)
  with that in a uniform electric field (red and thin). The two dotted lines indicate the oscillation amplitudes. The photodetachment spectra near a repulsive center
  is calculated and presented by using $d=200a_0$ and $\alpha=1$.
  The uniform field strength is selected as $128.55kV/cm$ which is equal to $\frac{\alpha}{d^2}$ in atomic units.
  The bottom abscissa shows the ticks for the photon energy as usual, and the corresponding scaled
  energy values are given at the top of the figure.}
\end{figure}

\begin{figure}[H]
\centering
  \includegraphics[width=300pt]{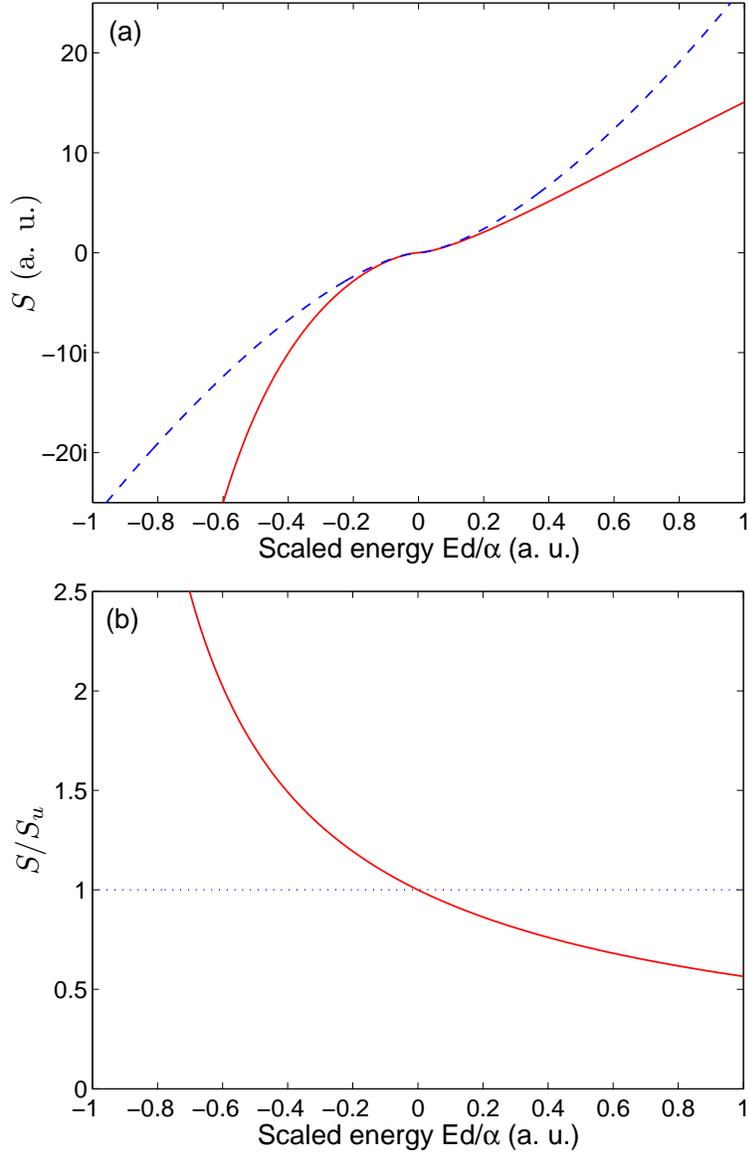}
  \caption{(Color online) Phase comparison between the photodetachment near a repulsive center and that in a uniform electric field.
  (a) The action $S$ accumulated by the detached electron after moving along the closed orbit near a repulsive center (solid curve with $d=200a_0$ and $\alpha=1$)
  is compared with that in a uniform field (dashed line with $F=128.55kV/cm$).
  Note the action value in the negative scaled energy range is pure imaginary.
  (b) The ratio of $S$ and $S_{u}$ is displayed and is only dependent on the scaled energy (see the context).
  The dotted line denotes the location of unity.
  Note in subplot (b) the solid curve also presents the ratio between $\frac{A}{kR}$ and $\frac{1}{3S}$
  for the photodetachment near a repulsive center.}
\end{figure}

\begin{figure}[H]
\centering
  \includegraphics[width=400pt]{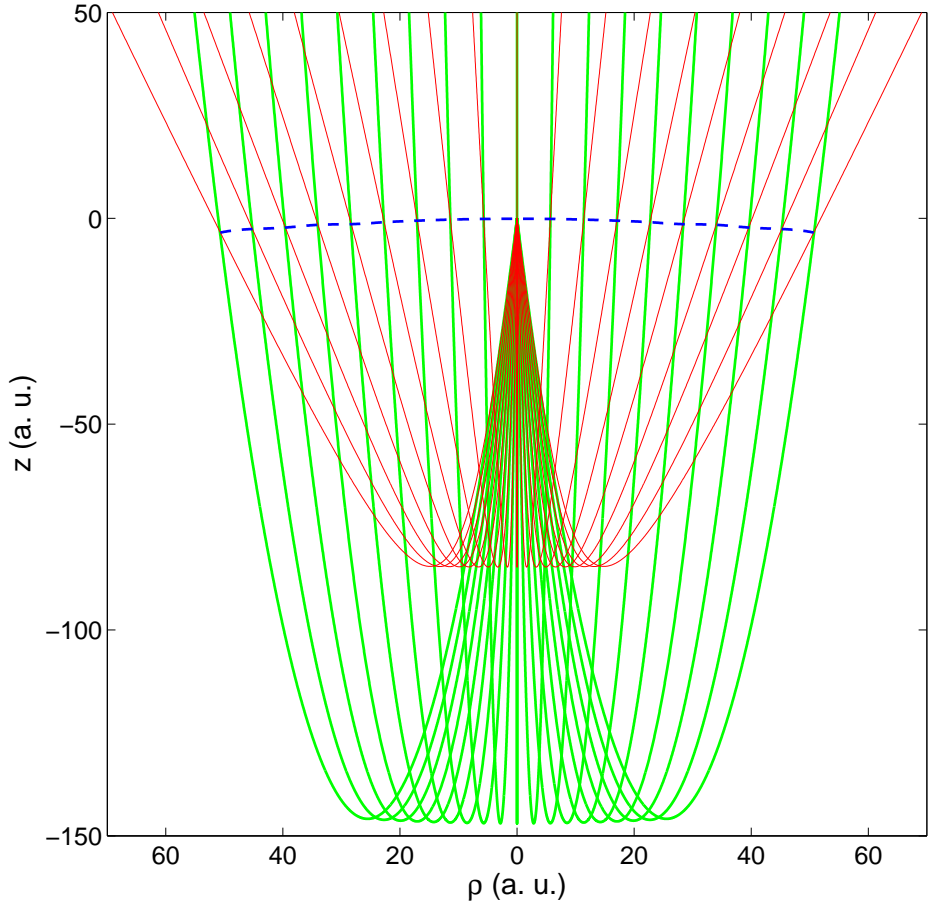}
  \caption{(Color online) Families of trajectories that illustrate why amplitudes of oscillation are equal for both the photodetachment processes near a repulsive Coulomb field
  and in a corresponding effective electric field. The thin curves (red) and the thick curves (green) depict, respectively,
  a pencil of trajectories adjacent to the closed orbit in a repulsive Coulomb field and that in the corresponding
homogenous static field, where we used  $d = 200a_0$, $\alpha = 1$ and $E=0.1eV$. The range of the initial ejection angles is from zero to five
degree relative to the closed orbit (note the different scales of horizontal and vertical axes). The dashed curve indicates a curve on which the two
trajectories (red vs. green online) starting at the same ejection angle cross each other. On this curve, the flux density through surfaces
perpendicular to the trajectories must be equal. At the source, those surfaces coincide, and are perpendicular to the closed orbit, therefore the
flux densities of returning waves for the red and green (online) families are equal at the closed orbit, and therefore the magnitudes of interference
oscillations are equal.}
\end{figure}

\end{document}